
\tolerance=2000
\magnification=1200
\hsize=6.0 truein
\vsize = 9.0 truein
\baselineskip=18 pt
\def\fprime{f^\prime}
\def\Sagb{{S_{\alpha\gamma}}^\beta}

\def\Qagb{{Q_{\alpha}}^{\gamma\beta}}
\def\halb{{1 \over 2}}
\centerline{TORSION AND NONMETRICITY IN SCALAR-TENSOR THEORIES OF GRAVITY}
\vskip 1.0 truein
\centerline{BY}
\vskip 1.0 truein
\centerline{\bf Jean-Paul Berthias and Bahman Shahid-Saless}
\centerline{Jet Propulsion Laboratory}
\centerline{301-150}
\centerline{California Institute of Technology}
\centerline{Pasadena, CA 91109}
\vskip 2.0 truein
\centerline{ABSTRACT}
\vskip 0.2 truein
\noindent
We show that the gravitational field equations derived from an action composed
of i) an arbitrary function of the scalar curvature and other scalar fields
plus ii) connection-independent kinetic and source terms, are
identical whether one chooses nonmetricity to vanish and have non-zero
torsion or vice versa.
\vfil
\eject

Scalar-Tensor theories of gravitation have been around for decades. These
theories, commonly based on actions representing coupling between the
gravitational metric tensor and other scalar fields have been of interest
for various reasons since the birth of General Relativity (GR).$^1$ The most
well know example is the Brans-Dicke theory$^2$ which was proposed
to incorporate Mach's principle in gravitational interactions.
Dirac$^3$ based his large-number hypothesis on
a scalar-tensor theory of gravity. Other theorists favored such theories
simply because of the presence of a scalar field which seems to be an
inevitable bi-product of modern physics. More recent examples of such models
include conformal-invariant theories of gravity, low-energy limit of
superstring
theories, Einstein-Cartan type theories coupled to scalar fields and many
modern inflationary models based on scalar-tensor theories.

In deriving the field equations of a theory of gravity from an action
functional
one is faced with choices. The most common approach
is to consider the metric tensor as
the only independent field describing the geometry of space-time and to
restrict the affine connections to be the well known Christoffel symbols.
This guarantees the metricity of the theory.
Thus Riemannian structure and local Lorentz structure are
preserved and the metric
is the solution to the ten metric field equations derived from the variation
of the action.  Another choice is to follow the Palatini formalism which
is to
consider the metric and the connections as independent fields. This choice
allows the geometry to have a general affine structure. The space-time
associated with this type of theory is usually called $(L_4,g)$. Here
there are in general
10 equations for the metric tensor and 64 for the connections.
In many cases this increase
in the number of equations is compensated by the reduction in order
of the metric field equations. For example in quadratic gravity the Palatini
variation yields two sets of first order equations for the metric and the
connections whereas the usual metric
formalism yields fourth order equations for the metric.
In the case of general
relativity the two formalisms produce identical field equations.

The advantage of deriving the field equations using the Palatini method
is that the geometry of space-time is less restrictive. In general the Palatini
variation allows for the existence of torsion and non-metricity.
The possible
importance of such exotic fields was not realized until the
construction of a gauge theory of gravitation was attempted. In the more
recent
years the inclusion of such fields in gravitational interactions and the study
of their properties have become more common as attempts to
unify all fundamental interactions with standard GR have failed in one
way or another.
The simplest theory incorporating torsion and nonmetricity is the
Einstein-Cartan-Sciama-Kibble (ECSK) theory$^4$ in which torsion is coupled
to the canonical spin tensor via the matter part of the lagrangian. In this
theory torsion does not propagate and also vanishes in vacuum. The Brans-Dicke
version of the ECSK theory (ECSK minimally coupled to a scalar field),
with non-zero torsion and vanishing non-metricity, which we denote by BDT,
was discussed by Rauch$^5$, German$^6$, Kim$^7$ and others. It was shown
that the scalar
field can act as a source of propagating torsion even in vacuum.
Furthermore Smalley$^8$ showed that the simplest non-metric version of the
Brans-Dicke
model, with vanishing torsion, which we denote by BDN here,
is equivalent to the BDT theory via a conformal transformation.

In this paper we generalize Smalley's result to include a much larger
class of scalar-tensor theories. Furthermore, we do this by considering
the so-called projective transformation of the connections rather than by
extended
conformal transformations. This can be done because we allow nonmetricity
to exist. We prove that given any scalar-tensor action
composed of any arbitrary function of the scalar curvature and other scalar
fields plus any kinetic and/or matter terms, which are
independent of the connections,
the field equations are independent of whether the torsion is set to zero
and non-metricity is nonzero or non-metricity is set to zero and the torsion
is non-zero. We show that this equivalence is a result of the
projective gauge
invariance of the action. We prove our results
by showing that: a) two distinct choices of the projective gauge vector
correspond
to the two cases mentioned above, and b) the two cases considered
in a) result in identical field equations.
Our results then imply that the equivalence between
the BDT and BDN theories is a consequence of this gauge freedom.
As a further example we apply our results to the case of
another simple action which involves a quadratic term
in the scalar curvature.

Our index conventions are the same
as those of Held et. al.$^9$ The connections are defined such that when a
vector
$A^\lambda$ is parallel transported, it undergoes an infinitesimal change
given by:
$$ d A^\lambda = - {\Gamma^\lambda}_{\mu\nu}(x) A^\nu dx^\mu.$$

\vskip .2 truein
\centerline{\bf THE FIELD EQUATIONS}
\vskip .2 truein

Let us consider the action:

$$ A = \int \lbrack f(\phi, R){\sqrt -g}  + L_{matter} + L_{Kinetic} \rbrack
d^4 x, \eqno(1)$$

\noindent
where $L_{matter}$ and $L_{Kinetic}$ represent the matter part of the
lagrangian and the kinetic energy terms associated with the fields
respectively. These are assumed to be independent of the connection fields.
The only term depending on the connections is the function $f(\phi, R)$
through its dependence on the scalar curvature $R(g,
{\Gamma})=g^{\mu\nu}R_{\nu\mu}({\Gamma})$ where $g$
and ${\Gamma}$ denote the metric and the connection fields. In this paper we
are
primarily interested in the field equations for the connections. Therefore
the analytic form of the kinetic and the matter terms and their dependence
on other fields are of no importance here.

Variation of the action Eq. (1), with
respect to the connections ${\Gamma}{^\alpha}_{\mu
\nu}$ gives:

$$\eqalign{{S_{\alpha \gamma}}^\beta +
2 {\delta^\beta}_{[\alpha} S_{{\gamma}]} +
{\delta^\beta}_{[\alpha} Q_{\mu]}{^\mu}{_\gamma} - 2 {\delta^\beta}_{[\alpha}
Q_{{\gamma}]} =&
- {\delta^\beta}_{[\alpha} \partial_{{\gamma}]} {\rm ln} \fprime
\cr
\equiv& {{P_{\alpha\gamma}}^\beta}, }\eqno(2)$$

\noindent
where

$$\eqalign{\Sagb \equiv& {\Gamma^\beta}_{[\alpha\gamma]}
\cr
=& \halb({\Gamma^\beta}_{\alpha\gamma} - {\Gamma^\beta}_{\gamma\alpha})}
\eqno(3)$$

\noindent
is the torsion tensor and

$$\Qagb = \nabla_\alpha g^{\gamma\beta}, \eqno(4)$$

\noindent
is the non-metricity tensor. The square brackets denote anti-symmetrization
as in Eq. (3) and the prime on $f$ denotes partial differentiation with
respect to $R$. Also in Eq. (2) $Q_\gamma \equiv {1\over 4}
{Q_{\gamma\beta}}^\beta$.

Multiplying Eq. (2) by $g^{\alpha\gamma}$ we get:

$${\delta^\beta}_{[\gamma} {Q_{\mu]}}^{\mu\alpha} = 0,\eqno(5)$$

\noindent
which implies that four of the sixty four equations do not contain any
information. This implies that the connections can only be determined to
within an arbitrary four-vector $V^\alpha$.
We will show that the freedom of choosing $V^\alpha$
corresponds to the well known projective invariance
of the action discussed by other authors$^{10}$.

In order to construct a unique theory of
scalar-tensor gravity one is usually forced to choose this vector. In torsion
theories of gravity, the choice of setting the non-metricity equal to zero
corresponds to taking the projective freedom away and Eq. (5)
is trivially satisfied. In this way the connections can be uniquely determined
in terms of the Christoffel symbols and the torsion tensor.

Let us define the vector

$$V^\alpha = {Q_\beta}^{\beta\alpha}.\eqno(6)$$

\noindent
Substitution of Eq. (6) into the field equations Eq. (2) gives:
$$S_{\alpha\beta\gamma} - \halb Q_{\alpha\beta\gamma}=P_{\alpha\beta\gamma}
+ \halb g_{\gamma\alpha} V_\beta - g_{\gamma\beta} V_\alpha - g_{\gamma\alpha}
S_\beta + g_{\gamma\beta}S_\alpha, \eqno(7)$$

\noindent
where $S_\alpha \equiv {S_{\alpha\beta}}^\beta $ and $P_{\alpha\beta\gamma}=
 -g_{\gamma[\alpha} \partial_{\beta]} {\rm ln} \fprime.$

The connections are given by$^{9}$:

$$\eqalign{ {\Gamma^{\sigma}}_{\alpha\beta} \equiv  g^{\sigma\mu}
\Delta^{\eta\lambda\tau}_{\beta\alpha\mu}\left(\halb \partial_\eta
g_{\lambda\tau} - S_{\eta\lambda\tau} + \halb
Q_{\eta\lambda\tau}\right)},\eqno(8a)$$

\noindent
where $\Delta^{\eta\lambda\tau}_{\beta\alpha\mu}$ is the permutation tensor
given by:

$$\Delta^{\eta\lambda\tau}_{\beta\alpha\mu}={\delta^\eta}_\beta
{\delta^\lambda}_\alpha {\delta^\tau}_\mu +
{\delta^\eta}_\alpha {\delta^\lambda}_\mu {\delta^\tau}_\beta
-{\delta^\eta}_\mu {\delta^\lambda}_\beta {\delta^\tau}_\alpha.$$

\noindent
Substitution of Eq. (7) in Eq. (8a) gives:

$$\eqalign{ {\Gamma^{\sigma}}_{\alpha\beta}= &
\left\{ \sigma\atop{\alpha\beta} \right\}
- g^{\sigma\mu} \lbrack P_{\beta\alpha\mu} + P_{\alpha\mu\beta} -
P_{\mu\beta\alpha} \rbrack  \cr
& - g^{\sigma\mu} \left\{
 -  \bigl[ {3\over2}g_{\alpha\beta}V_\mu - \halb g_{\beta\mu}V_\alpha -
{3\over2}g_{\alpha\mu}V_\beta \bigr] + 2 \bigl[ g_{\alpha\beta} S_\mu
-g_{\alpha\mu} S_\beta \bigr] \right\},}\eqno(8b)$$

\noindent
where $\left\{\sigma\atop{\alpha\beta}\right\}$ are the Christoffel symbols.

The torsion tensor can now be
found by antisymmetrizing the connections given by Eq. (8b). We
get:

$$\eqalign{{S_{\alpha\beta}}^\gamma = & {\Gamma^\gamma}_{[\alpha\beta]} \cr
= &\halb {\delta^\gamma}_{[\alpha} \partial_{\beta]}
{\rm ln} \fprime - \halb {\delta^\gamma}_{[\alpha} V_{\beta]}.} \eqno(9)$$

\noindent
Substituting this back into Eq. (8b) we get:

$$ {\Gamma^{\sigma}}_{\alpha\beta} =
\left\{ \sigma\atop{\alpha\beta} \right\}
+ g^{\sigma\mu} g_{\alpha[\beta} \partial_{\mu]} {\rm ln} \fprime - \halb
{\delta^\sigma}_\beta V_\alpha. \eqno(10)$$

\noindent
The non-metricity is found by taking the covariant derivative
of the metric using the full connections Eq. (10). After some algebra we
get:

$$\eqalign{ {\nabla_\alpha} g^{\beta\gamma} = &{Q_\alpha}^{\beta\gamma} \cr
= & V_\alpha g^{\beta\gamma}.}\eqno(11)$$

\noindent
Note that Eq. (10) is not an explicit solution for the connections but rather
an implicit equation since the connections are present in the
function $f$ itself.

\noindent
The Ricci tensor can now be calculated using the standard definition. We
have:
$$\eqalign{ R_{\beta\gamma} \equiv& {R_{\sigma\beta\gamma}}^\sigma \cr
=& {\Gamma^\sigma}_{\beta\gamma,\sigma} - {\Gamma^\sigma}_{\sigma\gamma,\beta}
+ {\Gamma^\sigma}_{\sigma\lambda} {\Gamma^\lambda}_{\beta\gamma}
- {\Gamma^\sigma}_{\beta\lambda} {\Gamma^\lambda}_{\sigma\gamma} \cr
=& R_{\beta\gamma}\left(\left\{\right\}\right) + \halb g_{\beta\gamma}
D_\sigma \partial^\sigma {\rm ln} \fprime
+ D_\beta \partial_\gamma {\rm ln} \fprime
-\halb g_{\beta\gamma} (\partial_\sigma {\rm ln} \fprime)( \partial^\sigma {\rm
ln} \fprime)
\cr
&+\halb
\left(\partial_\beta {\rm ln} \fprime\right)
\left(\partial_\gamma {\rm ln} \fprime\right) + D_{[\beta}
V_{\gamma]},}\eqno(12)$$

\noindent
where $R_{\beta\gamma}\left(\left\{\right\}\right)$ and $D_\alpha$ are the
Ricci tensor and the covariant derivative derived from the Christoffel symbols.

Inspection  of Eq. (12) reveals the independence of the scalar curvature and
thus the action on the vector $V_\alpha$ since $D_{[\beta}
V_{\gamma]}g^{\beta\gamma} = 0$ for a symmetric metric. The vector
$V_\alpha$ is the projective gauge vector.
Furthermore we note that if $V_\alpha$ is the derivative of a scalar
one can show that the
Riemann-Cartan tensor, the Ricci tensor itself and therefore the metric field
equations are also
independent of $V_\alpha$.  Two cases are of interest
here:

{\bf Case A}: $V_\alpha = 0$;
Here non-metricity vanishes via Eq. (9). and the torsion field is given by:

$${S_{\alpha\beta}}^\gamma =
\halb {\delta^\gamma}_{[\alpha} \partial_{\beta]}
{\rm ln} \fprime  \eqno(13)$$

\noindent
and

$$\eqalign{R_{\beta\gamma}=&
R_{\beta\gamma}\left(\left\{\right\}\right) + \halb g_{\beta\gamma}
D_\sigma \partial^\sigma {\rm ln} \fprime
+ D_\beta \partial_\gamma {\rm ln} \fprime
-\halb g_{\beta\gamma} (\partial_\sigma {\rm ln} \fprime)( \partial^\sigma {\rm
ln} \fprime)
\cr
&+\halb
\left(\partial_\beta {\rm ln} \fprime\right)
\left(\partial_\gamma {\rm ln} \fprime\right). }\eqno(14)$$

{\bf Case B}: $V_\alpha = \partial_\alpha {\rm ln} \fprime$;
In this case torsion vanishes via Eq. (10) and non-metricity is given by:

$${Q_\alpha}^{\mu\nu} = g^{\mu\nu} \partial_\alpha {\rm ln} \fprime,
\eqno(15)$$

\noindent
and $R_{\mu\nu}$ remains the same as in Eq. (14) because $V_\alpha$ is a
derivative of a scalar. Therefore the field equations in the two cases are
the same.

For the action corresponding to the Brans-Dicke theory, $f(\phi, R)= \phi
R$. Case A corresponds to the BDT theory discussed by Refs. (5-7).
Case B corresponds to the BDN theory discussed by Smalley.$^8$ It
is clear that the two theories are equivalent.
It is interesting to note that an extended conformal transformation which would
gauge away the torsion field for the BDT theory was found by German$^6$ which
corresponds
to the choice of $V_\alpha$ made in case B.  However,  the freedom of having
nonmetricity
here makes it unnecessary to rescale the metric,  or even the scalar field,  to
make the
action invariant.

\vskip .2 truein
\centerline{\bf EXAMPLE}
\vskip .2 truein

As a further example we consider another
class of actions, which has regained its popularity
in recent years, because of its emergence in
the low energy limit of superstrings.$^{11}$
The simplest such action is given in terms
of the Hilbert-Einstein term plus a quadratic scalar term, $f(R)=R+\alpha
R^2$. Recently the
non-metric version of this theory has been discussed by
Shahid-Saless.$^{12-14}$ It
was shown that, assuming no torsion,
the Palatini variation of this action yields a non-metric
theory. A more general version of this type of theory which assumes vanishing
non-metricity but includes torsion
was discussed by many authors.
In particular Minkevich$^{15}$ derived and examined the cosmological
field equations based on an action which included all the possible quadratic
combinations of curvature.
Given the formalism developed here it is clear that the
two theories yield the same field equations in the limit that the action
considered by Minkevich
corresponds to that examined by Shahid-Saless; that is $f(R)= R +
\alpha R^2$. In the case of vanishing torsion Eqs. (10) and (9) imply:

$$\eqalign{{Q_\beta}^{\mu\nu} =&
g^{\mu\nu} {2\alpha R_{,\beta} \over 1+ 2\alpha R}\cr
=& g^{\mu\nu} V_\beta}$$

\noindent
and the connections are:

$$ {\Gamma^{\sigma}}_{\alpha\beta} =
\left\{ \sigma\atop{\alpha\beta} \right\}
+{\alpha \over 1+2\alpha R}\left({\delta^\sigma}_\beta R_{,\alpha} +
{\delta^\sigma}_\alpha R_{,\beta} - g^{\sigma\mu}g_{\alpha\beta}
R_{,\mu}\right),\eqno(16)$$

\noindent
which agree with the results given by Shahid-Saless. However since $V_\alpha$
is a total derivative, it does not contribute to the the Ricci tensor. Thus
Eq. (12) will agree with the expression for the Ricci tensor used by
Ref. (11).  Therefore even if we had set $V_\alpha=0$, we
would have the same expression for the Ricci tensor. This case would however
correspond to the theory considered by Minkevich which is a metric theory
with torsion. The equivalence of his theory with that considered by
Shahid-Saless can be inspected easily
by setting Minkevich's $f_6$ by $-\alpha$. The cosmological equations discussed
by these authors$^{13,15}$ have also been checked for their equivalence.

\vskip .2 truein
\centerline{\bf CONCLUSION}
\vskip .2 truein

We have proven that for all actions composed of i) an arbitrary function of
the scalar curvature and other scalar fields, plus ii) any other kinetic
and matter terms which are independent of the connections, the field equations
are the same whether the torsion is set to zero and the theory is non-metric
or non-metricity is set to zero but torsion is non-vanishing.

Traditionally non-metricity has been viewed as an unwanted bi-product of
some extensions of GR because of its volume non-preserving property. On
the other hand it is generally argued that the existence of torsion in nature
does not pose a problem for fundamental physics. Our results show that within
the class of scalar-tensor theories considered here the two fields result
in identical field equations and therefore imply the same physics. One
conclusion that could be made is that perhaps one needs a deeper understanding
of the meaning of these fields and their inter-relationships in all aspects
of measurement before discarding them as physically unreasonable mathematical
artifacts.

\vskip .2 truein
\centerline{\bf ACKNOWLEDGEMENT}
\vskip .2 truein

This work was performed at the Jet Propulsion Laboratory which is under
contract with
the National Aeronautics and Space Adiministration. BSS  was supported by a
resident
research associateship award from the National Research Council of the National
Academy
of Science.
\vfil
\eject
\centerline{\bf REFERENCES}
\vskip .2 truein

\item{1.}  A. Einstein, Sitz. Pruess. Akad. Wiss. Berlin, {\bf 47},
pp. 778, 799, 844 (1915).
\item{2.}  C. Brans and R. H. Dicke, Phys. Rev. {\bf 124}, 925 (1961) .
\item{3.}  P. A. M. Dirac, Proc. Roy. Soc. Lon. {\bf A333}, 403 (1973).
\item{4.}  D. W. Sciama, Rev. Mod. Phys. {\bf 36}, 463 and 1103 (1964).
\item{5.}  R. T. Rauch, Phys. Rev. Lett. {\bf 52}, 1843 (1984).
\item{6.}  G. German, Phys. Rev. D {\bf 32}, 3307 (1985).
\item{7.}  Sung-Won Kim, Phys. Rev. D {\bf 34}, 1011 (1986).
\item{8.}  L. L. Smalley, Phys. Rev. D {\bf 33}, 3590 (1986).
\item{9.}  F. Hehl, P. von der Heyde and G. D. Kerlick, Rev. Mod. Phys.
{\bf 48}, 393 (1976).
\item{10.}  See for example V. D. Sandberg, Phys. Rev. D {\bf 12}, 3013
(1975) and references therein.
\item{11.}  M. B. Green and J. H. Schwarz, Phys. Lett.  B{\bf 149}, 117 (1984);
M. B. Green and J. H. Schwarz, Nucl. Phys. B{\bf 255}, 93 (1985).
\item{12.}  B. Shahid-Saless, Phys. Rev. D {\bf 35}, 467 (1987).
\item{13.}  B. Shahid-Saless, J. Math. Phys. {\bf 31}, 2429 (1990).
\item{14.}  B. Shahid-Saless, J. Math. Phys. {\bf 32} (3), 694 (1991).
\item{15.}  A. V. Minkevich, Phys. Lett. A {\bf 80}, 232 (1980).

\end